# Efficient, Spectrally Tunable Single-Photon Sources Based on Chlorine-Doped ZnSe Nanopillars

Y. Kutovyi [a*], M.M. Jansen [a], S. Qiao [a], C. Falter [a], N. von den Driesch [a], T. Brazda [a], N. Demarina [b], S. Trellenkamp [c], B. Bennemann [a], D. Grützmacher [a], and A. Pawlis [a*]

[a] Peter Grünberg Institut (PGI-9) & JARA-FIT, Forschungszentrum Jülich GmbH, Germany

[b] Peter Grünberg Institut (PGI-2), Forschungszentrum Jülich GmbH, Germany

[c] Helmholtz Nano Facility (HNF), Forschungszentrum Jülich GmbH, Germany

*Email: a.pawlis@fz-juelich.de, y.kutovyi@fz-juelich.de

ABSTRACT. Isolated impurity states in epitaxially grown semiconductor systems possess important radiative features such as distinct wavelength emission with a very short radiative lifetime and low inhomogeneous broadening which makes them promising for the generation of indistinguishable single photons. In this study, we investigate chlorine-doped ZnSe/ZnMgSe quantum well (QW) nanopillar (NP) structures as a highly efficient solid-state single-photon source operating at cryogenic temperatures. We show that single photons are generated due to the radiative recombination of excitons bound to neutral Cl atoms in ZnSe QW and the energy of the emitted photon can be tuned from about 2.85 down to 2.82 eV with ZnSe well width increase from 2.7 to 4.7 nm. Following the developed advanced technology we fabricate NPs with a diameter of about 250 nm using a combination of dry and wet-chemical etching of epitaxially grown ZnSe/ZnMgSe QW well structures. The remaining resist mask serves as a spherical- or cylindrical-shaped solid immersion lens on top of NPs and leads to the emission intensity enhancement by up to an order of magnitude in comparison to the pillars without any lenses. NPs with spherical-shaped lenses show the highest emission intensity values. The clear photon-antibunching effect is confirmed by the measured value of the second-order correlation function at a zero time delay of 0.14. The developed single-photon sources are suitable for integration into scalable photonic circuits.

INTRODUCTION

Quantum emitters producing identical single photons on demand are a key element for the realization of photonic components in quantum information technologies.[1,2] Over the last two decades, significant research efforts have been devoted to develop stable and reliable single-photon sources (SPSs) showing highly efficient and scalable performance. In this regard, solid-state single-photon emitters based on isolated impurities hosted in epitaxially grown compound semiconductors are particularly interesting because of their ability to generate indistinguishable and high-purity single photons with a short radiative lifetime.[3,4] The emission of single photons in such devices is typically attributed to radiative recombination of excitons bound to spatially isolated shallow impurities (e.g. donors, acceptors).[4] Remarkably, due to the discrete nature of bound exciton energy states in such semiconductor systems, the single photons can be generated through the relevant optical transitions in a well-defined spatial and polarization mode (i.e. in the same quantum state), which is in the ideal case makes each photon naturally indistinguishable from all the others subsequently generated by the same device. Such a feature is a crucial prerequisite of many quantum phenomena for example two-photon interference and therefore is highly required for future applications of photonic quantum emitters.[5]

Zinc Selenide (ZnSe) has been intensively investigated over the last few decades as a promising II-VI semiconductor material with a direct bandgap possessing excellent properties for optical and electronic applications.[6–8] In this regard, ZnSe has emerged as a promising host crystal matrix for impurity-bound-exciton-related emission of single photons.[9,10] In particular, the ability to generate indistinguishable single photons was recently demonstrated with Fluorine (F) donors, spatially isolated in epitaxially grown ZnSe/ZnMgSe QW nanostructures (i.e. ZnSe:F).[3] Moreover, successful generation of substantial polarization entanglement between photons generated by independent F-impurity-based emitters was also recently reported[9] and comprises an essential step towards the realization of long-distance quantum communication and networking. Furthermore, the recent publications also showed that ZnSe:F provides isolated impurity-bound electrons which can be effectively used as optically addressable spin qubits.[10] Such features of ZnSe:F make it particularly attractive for modern quantum applications and therefore motivate further research of other impurities in epitaxially grown ZnSe.

More recently, Chlorine (Cl) impurities in ZnSe (i.e. ZnSe:Cl) were suggested as alternative impurity-based single-photon sources demonstrating stable and bright emission of single photons.[11] Similar to the F-impurity, Cl substitutes the Se atom in the ZnSe crystal yielding a shallow donor impurity. However, in contrast to F, Cl has an atomic radius that is comparable with that of the Se atom making Cl a perfect donor on the Se site that considerably better matches into the ZnSe crystal compared to F. At the same time, as shown in a recent investigation,[11] Cl impurities in ZnSe consolidate the advantages of F-donors to be exploited as reliable single-photon emitters. However, it should be noted that the related study[11] has been performed exclusively for unstructured Cl-doped ZnSe/ZnMgSe QW samples and didn't attempt to produce an independent device that can be integrated into a photonic quantum circuit.

In this work, we are going beyond existing research by presenting an advanced nanofabrication concept and report on the optical performance of independent Cl-bound quantum emitters developed and fabricated in form of ZnSe/ZnMgSe QW NP structures which are directly integrated with three-dimensional all-dielectric solid immersion lenses (SILs) to improve the external quantum efficiency of the devices. The employment of SILs for such purposes has been extensively discussed and shown in the literature over the last few decades for a variety of emitters.[12,13] It has been revealed that SILs are a suitable remedy to overcome the total internal reflection problem[12] which is the origin of the low collection efficiency of the emitted radiation coming out from the emissive material. For instance, the use of a SIL with a refractive index smaller than the refractive index of the semiconductor usually results in increased external quantum efficiency. With this in mind, we advanced a nanopillar fabrication technology approach[3] allowing us firstly to fabricate a SIL and then directly use it as a hard mask for the subsequent etching of the NP. It should be noted that in this case, deterministically fabricated nanolenses don't require any post-fabrication alignment and positioning as they become directly coupled with the NP emitters after the etching. We demonstrate that the developed approach allows for the fabrication of highly efficient quantum emitters with enhanced external quantum efficiency and high emission rates of antibunched photons. In addition, we address a common problem of spectral variability between photons generated by different devices. In this regard, we numerically and experimentally demonstrate that our deterministically engineered devices can effectively generate single photons with pre-defined emission energies which reveals the substantial potential of our ZnSe:Cl SPSs to form reliable optical quantum links in future quantum networks.

RESULTS AND DISCUSSION

**Device concept and single-photon emission properties.** Figure 1a shows a schematic illustration of the developed single-photon emitter which consists of a ZnSe/ZnMgSe QW NP structure coupled with a spherical-shaped HSQ nanolens. The operation principle of the device is based on the radiative recombination of excitons which are bound to a single spatially isolated Cl donor in an epitaxially grown ZnSe/ZnMgSe QW (see Section 1 in the Supporting information). A typical photoluminescence (PL) spectrum measured for an NP structure at a temperature of 5 K is shown in Figure 1b. The sharp line which we attribute to the emission of single photons due to the radiative decay of the Cl-bound exciton complex has been observed at about 2.848 eV. The origin of the peak is confirmed by the PL intensity of the line measured as a function of the excitation power for the NP device with a spherical-shaped HSQ lens at a temperature of 7 K (see Figure 2a). As can be seen in Figure 2a, the intensity of the narrow $D^0X$ peak exhibits a superlinear power-function dependence ($I \propto P^{1.13\pm0.05}$) in the low excitation regime (i.e. $P \leq 10$ μW). The dependence starts to saturate with a further increase of the excitation power which is typical for the impurity bound-exciton emission.[11] At the same time, a group of peaks associated with the emission from heavy-hole free-excitons is observed in the energy region higher than 2.860 eV. It should be noted that the presence of several peaks in the FX group can be attributed to monolayer fluctuations of the QW width in the sample resulting in digital variations of the excitonic transitions. As can be seen in Figure 1b, the energy separation between $D^0X$ and FX emission is about 12 meV in this specific case, which provides substantial spectral independence and stability.

Such a large separation results from the enhanced binding energy of the donor-bound excitons in the two-dimensional QW.

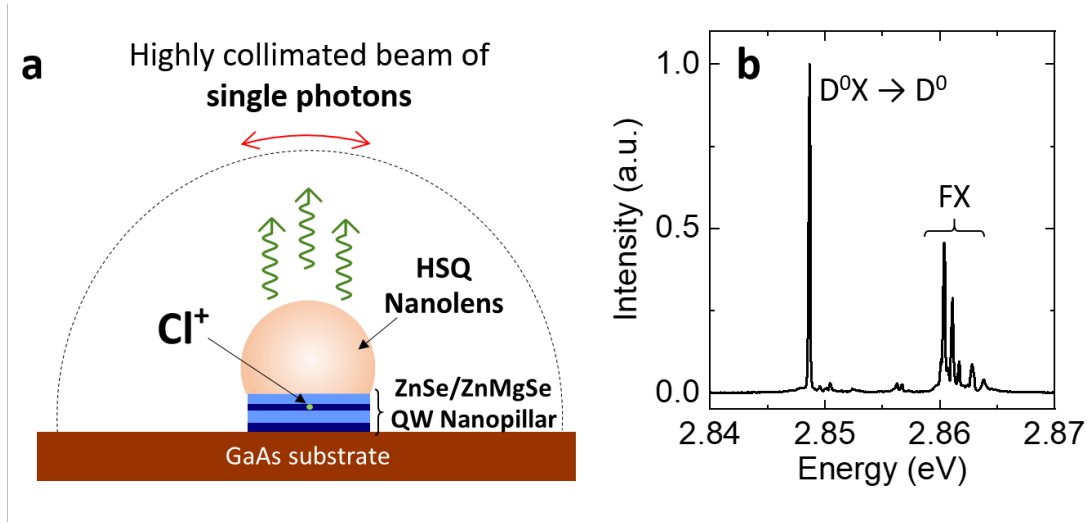

**Figure 1.** (a) Schematic illustration of a Cl-doped ZnSe/ZnMgSe QW NP with a spherical-shaped HSQ nanolens on top that helps to collect the emitted photons. (b) Typical PL spectrum measured for a ZnSe/ZnMgSe NP (ZnSe QW width of approximately 2.71 nm) with a spherical-shaped lens at a temperature of 5 K and with 5 μW pulsed laser excitation. The spectrum reveals a single-photon emission line at 2.848 eV that originates from the recombination of donor-bound excitons ($D^0X \rightarrow D^0$). The FX-labeled group of peaks corresponds to the recombination of heavy-hole free excitons.

To reveal that the fabricated Cl-doped ZnSe/ZnMgSe QW NPs are high-quality quantum emitters capable to emit single photons with high flux rates, we performed power-dependent PL measurements as well as time-correlated single-photon counting measurements.

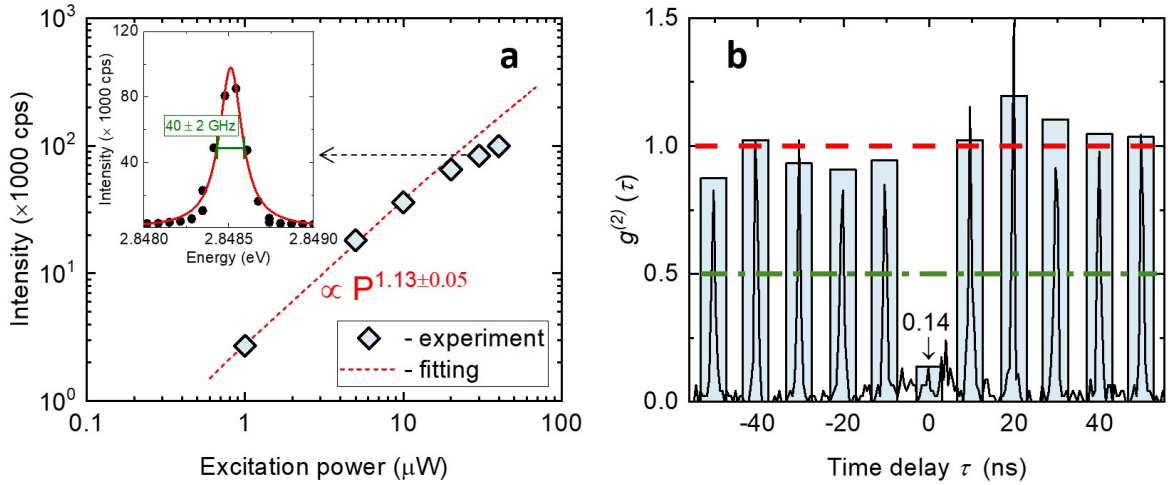

**Figure 2.** (a) Intensity of the $D^0X$ line (spectral shape shown in the inset) as a function of the excitation power measured at a temperature of 7 K. Dashed red line represents a linear fitting of the experimental data (in the log-log scale) measured at low-power excitations ($\leq 10$ µW). Inset: $D^0X$ peak of the PL spectrum measured at 30 µW as pointed out by the dashed arrow. (b) Normalized photon correlation histogram that was measured for the same device under 30 µW excitation power.

Figure 2b displays the normalized photon-correlation histogram measured for the donor-bound exciton emission from a typical NP emitter. The corresponding $D^0X$ emission line is shown in the inset of Figure 2a and was fitted with the Lorentz function yielding the spectral full-width-at-half-maximum (FWHM) of (40±2) GHz. It should be noted the FWHM of 40 GHz is limited by the resolving power of our spectrometer (see Methods) and is expected to be significantly smaller in the fabricated devices. At the same time, due to the high emission rate of the NP under investigation, an integration time of only 30 min was sufficient to achieve a clear coincidence correlation histogram. As can be seen in Figure 2b, the height of the central coincidence peak at zero delay time is considerably reduced revealing the clear photon-antibunching effect. In this exemplary case, we observed residual two-photon probability of $g^{(2)}(0)$ to be $0.14 \pm 0.02$ calculated from the ratio of the height of the central coincidence peak to the 20 averaged adjacent peaks ($\tau \neq 0$) and without any background subtraction. It should be noted that the obtained value of $g^{(2)}(0) = 0.14 \pm 0.02$ is well below the threshold value of 0.5 for quantum emitters.[3,11] We thus observe a clear single-photon antibunching emission from the single chlorine donor even at relatively high excitation powers (i.e. at 30 µW).

**Spectral tuning of quantum emission.** Spectral tuning is an essential aspect of the practical applications of single-photon sources in photon-based information processing schemes and readout protocols.[3,14] Note that the ability to fine-tune the PL emission and thus entangle single photons emitted from independent F donors in ZnSe QW nanostructures has been recently shown using a local laser heating method.[9] Below, we combine simulation and experiments to demonstrate the spectral coarse-tuning possibility for the developed single-photon emitters by engineering the ZnSe QWs with different widths.

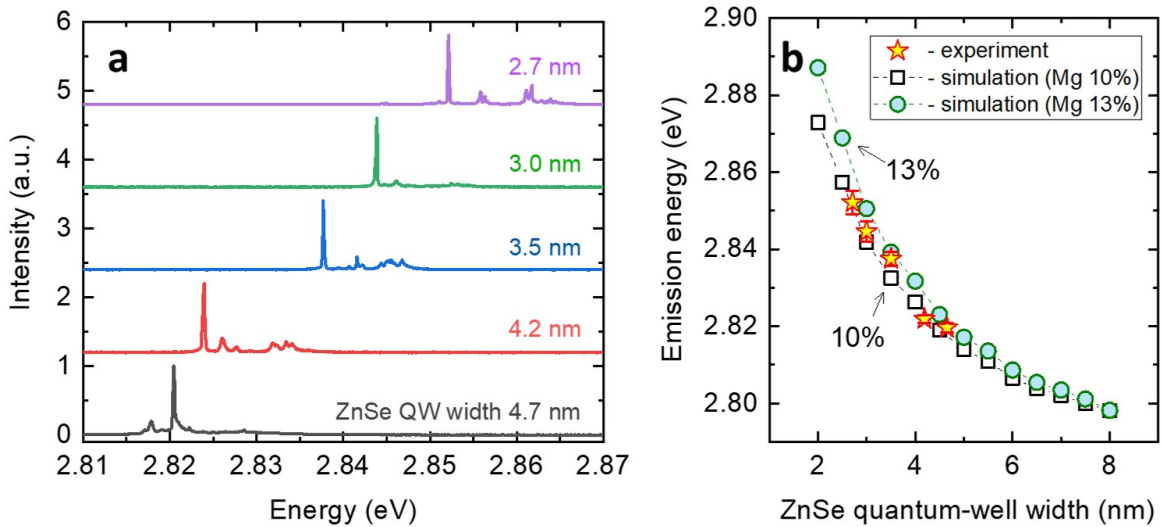

**Figure 3.** Coarse-tuning of single-photon emission. (a) Typical PL spectra measured for the NPs with different ZnSe QW widths under the same experimental conditions. The shift of the D⁰X

emission peak reflects the wavelength coarse-tuning of our developed NPs. (b) The emission energy of the donor-bound-exciton PL as a function of ZnSe QW width obtained for experimental and simulated data. The simulations were performed for the ZnMgSe/ZnSe/ZnMgSe QW stack with 10 % and 13 % Mg concentration as typical Mg contents in ZnMgSe barriers for the measured samples. Vertical error bars in the experimental data represent the standard deviation of the mean values calculated for each set of studied NPs with different QW widths.

Figure 3a displays typical PL emission spectra measured for chlorine-delta-doped NPs with different ZnSe QW widths. In this specific case, the PL measurements were performed at a temperature of 10 K and 5 µW excitation power using the NPs exhibiting a clear $D^0X$ emission related to the Cl-donor. As can be seen in Figure 3a, the emission energy of the bound-exciton PL shifts towards higher energy with decreasing QW width and allows for coarse-tuning of the emission wavelength. Figure 3b depicts the measured $D^0X$ peaks as a function of the ZnSe QW width. A pronounced redshift of the emission energy with increasing the QW width is observed. We attribute this redshift to the change of the QW confinement, also altering the exciton binding energy as a function of the QW width.[15] Moreover, as can be seen in Figure 3b, shrinking the QW width from 4.7 to 2.7 nm increases the emission energy of the $D^0X$ peak by 30 meV. This wide tuning range reveals the enormous effect of QW confinement on bound-exciton-related emission allowing us to precisely engineer and fabricate single-photon emitters with practically pre-defined emission properties.

To confirm our interpretation of the PL tuning experiments presented in Figure 3b, we calculated the donor-bound exciton ($D^0X$) binding energy in ZnSe/ZnMgSe QW structures with a theoretical model (see Methods) involving a nonvariational approach to density-functional theory[16,17] closely following the routine presented in Ref[17]. The calculated dependencies of the $D^0X$ emission energy as a function of ZnSe QW width are shown in Figure 3b. The calculations were performed for the ZnSe/ZnMgSe QW samples with 10 % and 13 % Mg concentrations as typical frame-values for the studied devices. As can be seen in Figure 3b, the obtained calculation results show an excellent agreement with the experimental data.

Finally, we present a new fabrication concept for the Cl-doped ZnSe/ZnMgSe NPs which are directly integrated with HSQ-based SILs (see Methods). Below we demonstrate and discuss the effect of SILs on top of the developed SPS comparing the optical performance of the NPs with spherical- and cylindrical-shaped HSQ nanolenses as well as without any lenses. The latter pillars were obtained after chemical etching with 1%-HF of the corresponding RIE-etched samples. Note that spherical-shaped lenses were intentionally defined by the performed grayscale electron-beam lithography and can be later further shaped using different etch chemistries (see Methods as well as Section 2 in the Supporting information for more details). Figures 4(a) and 4(b) displays the histogram plots of the $D^0X$ peak intensities measured for a set of pillars with and without HSQ lenses and processed with $H_2$/Ar and $H_2$/Ar/$CHF_3$ RIE plasmas, respectively. In this case, PL measurements for all pillars were performed under the same experimental conditions at a temperature of 5 K and 5 µW excitation power.

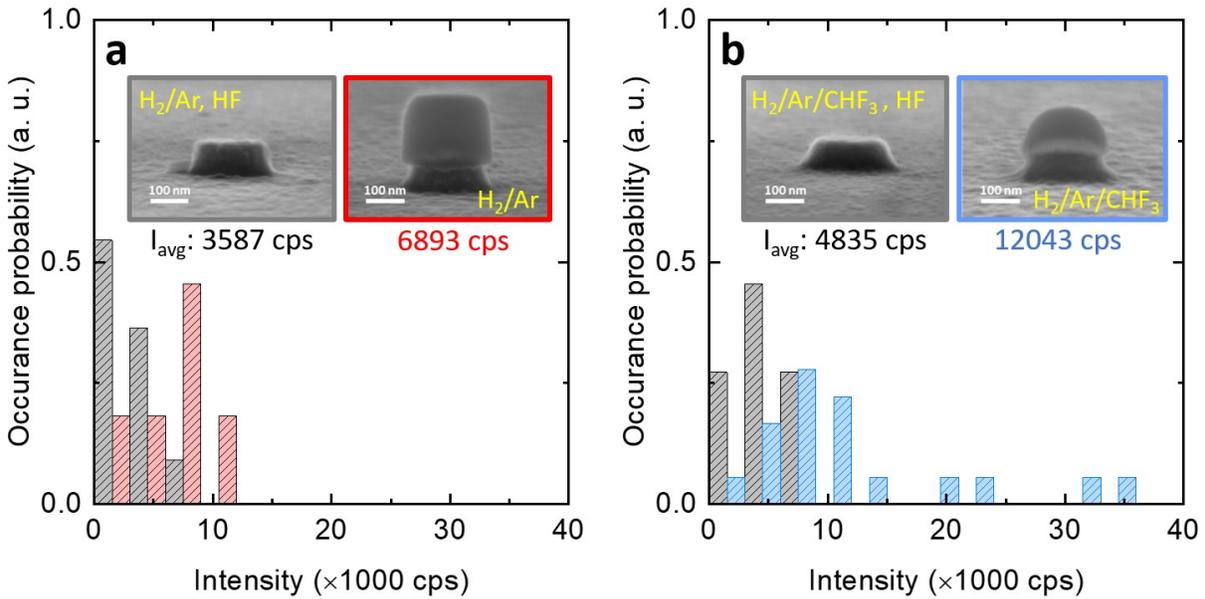

**Figure 4.** The histograms of the PL intensity distributions measured for the $D^0X$ emission from NPs with and without HSQ lenses etched using (a) $H_2/Ar$ or (b) $H_2/Ar/CHF_3$ RIE plasma. The histogram bin-width of 3000 cps is used. The mean average values of each intensity distribution are indicated under the corresponding SEM insets which represent a typical NP for each set of measured devices. For the NPs covered with spherical-shaped HSQ nanolenses and etched using $H_2/Ar/CHF_3$ RIE plasma (shown in blue), we observe up to one order of magnitude increase in the external quantum efficiency.

As can be seen in Figures 4(a) and 4(b), the measured intensity of the $D^0X$ emission for all studied emitters is statistically distributed with the mean values indicated in the plots. Intensity variations of about an order of magnitude are observed for all kinds of investigated devices regardless of the type and presence of the lens. Such behavior reflects a typical inhomogeneity issue for solid-state emitters[9,18] and can be attributed to lateral variations of donor position for different studied pillars. Our theoretical estimations using a geometrical ray-tracing approach presented in the Supporting information (see Section 3) reveal at least a factor of four for the intensity change between donor position in the center compared to that at the edge of the NPs covered with the spherical-shaped lenses. At the same time, for the devices with cylindrical-shaped lenses, a factor of about two was obtained. In this regard, on-demand single-dopant implantation technology[19] allowing for more precise positioning of individual donors in the NP would be the possible solution to reduce a device-to-device intensity variation related to the position of the optically active single donor (i.e. point-like source).

Despite the observed device inhomogeneity issue, a substantial improvement in the PL intensity for the pillars covered with cylindrical- and spherical-shaped HSQ lenses is well-pronounced and can be clearly evidenced in Figures 4(a) and 4(b), respectively. In particular, we see that the ratios between the average count rates for $H_2/Ar$ and $H_2/Ar/CHF_3$-etched pillars without and with lenses are about 1:1.75 and 1:2.49, respectively. Additionally, we see that the ratio between the average count rates for $H_2/Ar$ and $H_2/Ar/CHF_3$-etched pillars is about 1:1.35 when the lenses are removed with 1%-HF. This indicates a substantial positive impact of $CHF_3$ gas at low flow rates in the $H_2/Ar/CHF_3$ plasma on the internal quantum efficiency in the pillars due to the less-damaged side-wall surfaces compared to those obtained after dry etching with $H_2/Ar$ only.

Moreover, the effect of using HSQ as a mask and solid immersion lens is at least twofold: Firstly, by using HSQ material (i.e. glass-like $SiO_2$) with a low refractive index on top of the ZnSe/ZnMgSe NPs, we reduced the gradient in the refractive index change between semiconductor and environment, which leads to an increased total internal reflection angle and with that, better outcoupling efficiency.[12] This effect is also confirmed by numerical simulations performed using Lumerical simulation software. The simulation results are presented in the Supplementary information (see Section 4) and reveal that the presence of HSQ on top of ZnSe/ZnMgSe QW NPs offers up to 135% PL intensity enhancement, which is consistent with our experimental findings. This demonstrates substantial improvement in the outcoupling efficiency for NPs covered with HSQ.

Secondly, at the same time, NPs with spherical-shaped HSQ lenses generally demonstrate higher count rates compared to the devices covered with cylindrical-shaped HSQ. This partially reflects the effect of the spherical-shaped HSQ lens working as a plano-convex lens and effectively collimating the generated single photons into a smaller numerical aperture (see Figure S4c in the Supporting information).

CONCLUSION

Optically active impurities in epitaxially grown II-VI semiconductors offer a great opportunity to produce identical single photons on demand. In this study, we reported highly efficient single-photon emission as a result of the radiative recombination of excitons bound to single Cl donors which were intentionally isolated in ZnSe/ZnMgSe QW NPs. By two-photon correlation measurements, we verified single-photon antibunching for the devices hosting a single Cl-donor. A distinct spectral coarse-tuning up to 30 meV was experimentally and numerically demonstrated by engineering QWs with different widths. To improve the collection efficiency of single photons, we developed and fabricated NPs that are directly integrated with deterministic solid immersion nanolenses made of HSQ. As a result, a significant enhancement of the average external quantum efficiency by a factor of 2.5 for the devices with spherical-shaped HSQ nanolenses was achieved. These results pave the way for ZnSe:Cl-based single-photon sources as highly efficient and spectrally tunable devices to be applied in quantum computing and optically mediated quantum information processing technology.

# METHODS

**Sample fabrication.** The ZnSe/ZnMgSe QW nanopillars were fabricated on top of commercially available GaAs substrates with (100) crystallographic orientation. The details of the fabrication process flow can be found in the Supporting information (see Section 2). Briefly, epitaxial growth of Cl-doped ZnSe/ZnMgSe QWs was performed on GaAs substrates using molecular beam epitaxy (MBE). After MBE growth, the nanostructuring based on a top-down fabrication approach was carried out to pattern NP structures with SILs on top and thus isolate individual Cl impurities in ZnSe QWs. First, grayscale electron-beam lithography was employed to define nanolenses with a given size and shape in the HSQ resist. Next, the reactive ion etching (RIE) was performed to etch ZnSe/ZnMgSe NPs. Two recipes were developed and used for this purpose: In the first recipe, the ZnSe/ZnMgSe QW multilayer stack was etched using a gas mixture of hydrogen ($H_2$) and argon (Ar) leading to a cylindrical shape of the HSQ lenses. The second dry etching recipe is based on the mixture of $H_2$, Ar, and $CHF_3$ gases and was effectively used to pattern NPs with spherical-shaped HSQ nanolenses on top. The wet-chemical polishing using a water-based solution of potassium dichromate ($HBr:K_2Cr_2O_7:H_2O$) was then performed to remove any near-surface defects in the NPs introduced during the RIE step. Finally, to prevent oxidation and thus degradation of the nanostructures, all the samples were immediately passivated with 10 nm $Al_2O_3$ oxide deposited using an atomic layer deposition system.

**Optical characterization.** The nanostructures were excited non-resonantly at a wavelength of 394 nm using a pulsed frequency-doubled Ti:Sapphire laser with a pulse width of 80 fs working at a repetition frequency of 100 MHz. First, the emitted PL was collected using a high numerical aperture objective lens (NA = 0.9). Then, the emission was directed through a 1000 μm pinhole, dispersed with a 2400 lines holographic grating, and gathered to the CCD camera of a Princeton Instruments Acton SP 2500 monochromator with a 500 mm focal length. This system yields a spectral resolution of about 40 GHz at a 20 μm slit opening. For the two-photon correlation, the emission was filtered by the same monochromator system and transferred to a Hanbury-Brown-Twiss setup employing a 50/50 beam splitter and two single-photon sensitive avalanche photodiodes (Micro Photon Devices, PicoQuant) with timing resolution down to 50 ps. A time-correlated single-photon counting module (PicoHarp 300, PicoQuant) providing a channel resolution of 4 ps was used to perform the antibunching experiments.

**Calculation of the $D^0X$ binding energy in ZnSe/ZnMgSe QW structure.** The calculation of the $D^0X$ binding energy in ZnSe/ZnMgSe QW was performed considering the $D^0X$ complex consisting of a neutral donor, two electrons, and a hole (see Figure S1 in the Supporting information). The electron or hole effective Hamiltonian in the Kohn-Sham basis includes terms describing Coulomb interaction between the particles, the particles and the impurity ion as well as exchange-correlation interaction. During the sample growth, the ZnSe QWs are delta-doped in the middle thus in the calculations the Cl donors are assumed to be located at the center of the ZnSe QW. The form of the interaction-correlation potential is chosen following Refs[17,20]. The effective Hamiltonian for an electron and a hole can be decomposed into two parts: The first part represents the QW confinement while the second describes the motion in the plane perpendicular to the direction of the layer growth. The valence and conduction band discontinuities for ZnSe/ZnMgSe QWs are calculated taking into account the assumption that ZnSe and ZnMgSe layers are compressively strained being pseudomorphically grown on the GaAs substrate.[21] The electronic and structural parameters of ZnSe and ZnMgSe used for the calculations are taken from Refs[22,23].

Material parameters for $Zn_{1-x}Mg_xSe$ are obtained assuming that they change linearly with the Mg concentration. The calculated position of the conduction and valence bands and the component of the wave functions in the direction perpendicular to the layer growth for the Cl-doped ZnSe QW are shown in Figure S12 in the Supporting information. The slight bending of the bands is due to the electrical field associated with the ionized Cl atoms. To calculate the $D^0X$ binding energy, the ground-state energy of the neutral impurity as well as the ground-state energy of the free exciton[24] should be subtracted from the total energy of the system obtained as a result of solving the Kohn-Sham equation. Then the transition energy is calculated by subtracting the neutral impurity and $D^0X$ binding energies from the ZnSe band gap determined by the calculated electron and heavy hole ground-state energies.

AUTHOR INFORMATION

**Corresponding Author**

*Email: a.pawlis@fz-juelich.de, y.kutovyi@fz-juelich.de

**Author Contributions**

The manuscript was written with the contributions of all authors. All authors have approved the final version of the manuscript.

**Funding Sources**

This work is supported by Deutsche Forschungsgemeinschaft (DFG, German Research Foundation) under Germany's Excellence Strategy Cluster of Excellence: Matter and Light for Quantum Computing (ML4Q) (EXC 2004 1-390534769).

**Notes**

The authors declare no competing financial interest.

ACKNOWLEDGMENT

The authors acknowledge support from the technical staff of the Helmholtz Nano Facility (HNF) of Forschungszentrum Jülich for their assistance with the fabrication of the single-photon emitters. The authors thank Aziz Karasahin (the University of Maryland) for the help with the time-correlated single-photon counting measurements and discussion of antibunching experiments. This work is supported by Deutsche Forschungsgemeinschaft (DFG, German Research Foundation) under Germany's Excellence Strategy Cluster of Excellence: Matter and Light for Quantum Computing (ML4Q) (EXC 2004 1-390534769).

ABBREVIATIONS

QW, quantum well; NP, nanopillar; HSQ, hydrogen silsesquioxane; SPS, single-photon source; F, fluorine; Cl, chlorine; PL, photoluminescence; MBE, molecular beam epitaxy; SIL, solid immersion lens; RIE, reactive ion etching; HF, hydrofluoric acid; FWHM, full-width-at-half-maximum.

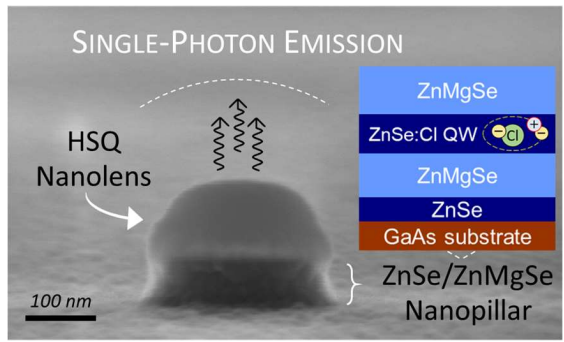

TOC image

# Supporting information for:

# Highly Efficient and Spectrally Tunable Single-Photon Sources Based on Chlorine-Doped ZnSe

Y. Kutovyi [a,*], M.M. Jansen [a], S. Qiao [a], C. Falter [a], N. von den Driesch [a], T. Brazda [a], N. Demarina [b], S. Trellenkamp [c], B. Bennemann [a], D. Grützmacher [a], and A. Pawlis [a,*]

[a] Peter Grünberg Institut (PGI-9) & JARA-FIT, Forschungszentrum Jülich GmbH, Germany

[b] Peter Grünberg Institut (PGI-2), Forschungszentrum Jülich GmbH, Germany

[c] Helmholtz Nano Facility (HNF), Forschungszentrum Jülich GmbH, Germany

*Email: a.pawlis@fz-juelich.de, y.kutovyi@fz-juelich.de

## 1. Working principle of the chlorine-donor-based single-photon emitter

The operation principle of the chlorine(Cl)-donor-based single-photon emitters is based on the radiative recombination of excitons which are bound to a spatially isolated Cl-donor in an epitaxially grown ZnSe/ZnMgSe quantum well (QW).[1,2] When ZnSe is above-bandgap excited with a laser source ($\lambda$ < 440 nm), free excitons (i.e. quasi-particles consisting of an electron and a hole bound via Coulomb interaction) are photogenerated and therefore might be eventually captured by a neutral Cl donor ($D^0$ state) forming a donor-bound exciton complex ($D^0X$ state). As schematically shown in Figure S1, the following recombination decay process of such a complex ($D^0X \rightarrow D^0$) yields an efficient emission of a single photon at a time, which makes this process an ideal mechanism for single-photon generation.

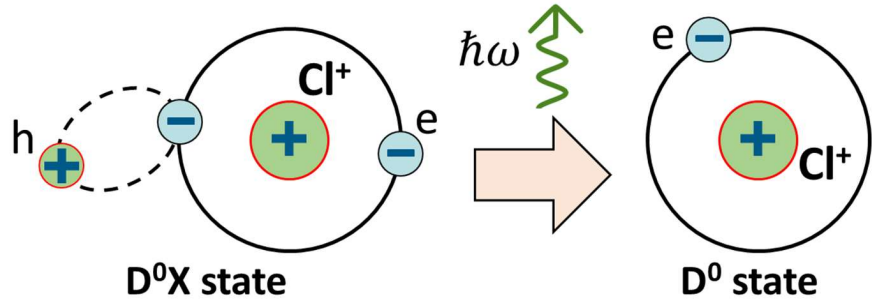

**Figure S2.** The recombination process of a Cl-donor-bound exciton complex in ZnSe resulting in the emission of a single photon (schematically).

## 2. Device fabrication process flow

In our study, we considered a common problem of photon collection efficiency[3] for the solid-state emitters and developed a new nanofabrication concept allowing us to fabricate Cl-doped ZnSe/ZnMgSe nanopillar (NP) emitters directly integrated with solid immersion lenses (SILs) which considerably improve the external quantum efficiency.[3,4] The developed concept is based on the deterministic fabrication of three-dimensional (3D) all-dielectric solid immersion lenses (SILs) which later served as a hard mask for the etching of NPs. Below, we describe the fabrication flow in detail.

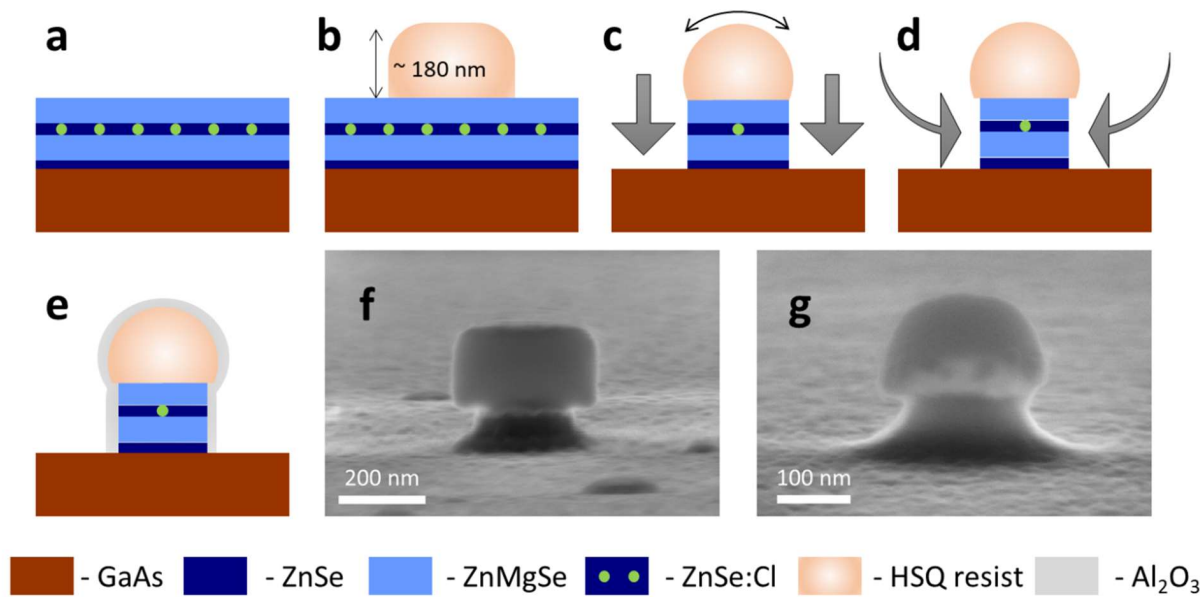

**Figure S2.** Process workflow for the fabrication of Cl-doped ZnSe/ZnMgSe QW NP structures with spherical-shaped solid immersion lenses using the top-down approach: (a) – MBE growth, (b) – e-beam lithography, (c) – reactive ion etching, (d) – wet-chemical polishing, (e) – passivation. (f) and (g) Typical SEM images of the ZnSe/ZnMgSeQW NPs with cylindrical-shaped and spherical-shaped nanolenses on top of the pillars, respectively.

The major steps of the developed fabrication technology are schematically depicted in Figure S2(a-e). As a first step, the epitaxial growth using molecular beam epitaxy (MBE) system was performed in ultra-high vacuum chambers for III-V and II-VI materials to grow Cl-doped ZnSe/ZnMgSe QWs on 2-inch GaAs wafers (see Figure S2a). For this purpose, the GaAs substrates were first deoxidized under As flux in the III-V chamber. Then, a few hundred nanometers thick GaAs buffer layer was grown to restore the atomically flat surface of the substrates. Afterward, the samples were transferred to the II-VI chamber under high-vacuum conditions, where a thin (5-15 nm) ZnSe buffer layer was grown below the subsequent growth of the ZnSe/ZnMgSe QW multilayer device stack to guarantee excellent heterointerface properties.

Typically, the QW stack consists of the Cl delta-doped ZnSe QW with a thickness of 1-5 nm and doping concentration of about $2\times10^{16}$ cm$^{-3}$ that is sandwiched between two (25-35 nm) thick ZnMgSe barrier layers with an Mg content of about 10-13%. The growth stoichiometry as well as the surface quality of the epitaxial layers have been probed directly after the growth using the reflective high energy electron diffraction method for each sample investigated in this study. Also, the strain status, uniformity, and thickness of the epilayers, as well as their chemical composition (e.g. Mg content), were measured using high-resolution X-ray diffraction.

After the MBE growth, the nanostructuring based on a top-down fabrication approach (see Figure S2c) was performed to define NP structures and thus isolate individual Cl impurities in ZnSe QWs. First, samples with ZnSe/ZnMgSeQW were treated with oxygen plasma for 3 min to remove SeO$_2$ clusters. After O$_2$ plasma treatment, a commercially available hydrogen silsesquioxane (HSQ) electron beam resist was immediately spin-coated on the samples. The Si-H bonds of the HSQ polymer break under e-beam exposure resulting in Si-O crosslinking which transforms the exposed areas into insoluble glass-like SiO$_2$ structures providing a refractive index that is lower than the refractive index of ZnSe/ZnMgSe. This allows us to increase the critical angle for the photons escaping from NPs making the HSQ-transformed silica a promising material for the fabrication of SILs to be used on top of ZnMgSe/ZnSe-based emitters. In our case, the spin-coating of the HSQ resist was performed at 4000 rpm for 30 s resulting in a film thickness of about 200 nm. The hard bake at 130 °C was further performed for 5 min to stabilize the HSQ polymer film. The SIL structures were then defined by grayscale e-beam lithography in the HSQ resist as follows.

To form nanolenses with a given size and shape in the HSQ resist, we defined concentric rings with a width of 60 nm centered at the given positions. The e-beam exposure was performed at room temperature at the acceleration voltage of 100 kV. The electron dose for the first central ring was set to 2200 µC/cm$^2$ to locally overexpose the resist and set the center point of the lens. Each following ring was written using an intentionally defocused electron beam with the electron dose reduced by a factor of 0.75. Such a gradient exposure approach allowed us to alter the local thickness of each exposed ring and thus precisely tailor a 3D shape of the lens. After the e-beam exposure, the structures were developed in MF CD 26 developer for 60 s and subsequently purged with de-ionized water (see Figure S2b). A typical SEM image of the spherical-shaped HSQ lens obtained after the resist development is shown in Figure S3.

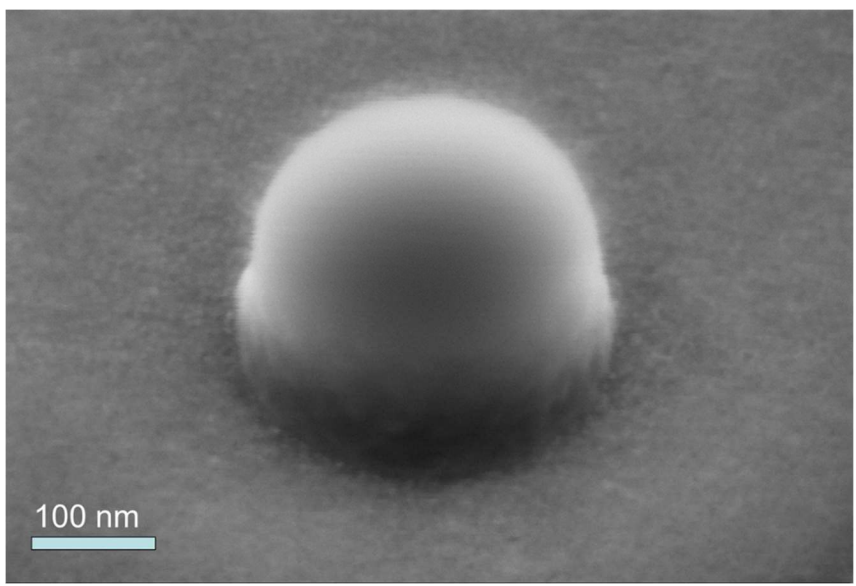

**Figure S3.** Scanning electron micrograph of a spherical-shaped HSQ nanolens on the ZnSe/ZnMgSe surface that was defined utilizing a grayscale e-beam lithography. The nanolens was used for the subsequent dry etching of the ZnSe/ZnMgSe QW NP.

As the next step, the reactive ion etching (RIE) technique was employed to etch ZnSe/ZnMgSe QW NPs (see Figure S2c). Two recipes were developed and used for this purpose. In the first recipe, the ZnSe/ZnMgSe multilayer stack was etched using a gas mixture of hydrogen (H$_2$) and argon (Ar) with flow rates of 60 and 20 standard cubic centimeters per minute (sccm), respectively. A typical SEM image of the NP device after H$_2$/Ar plasma etching is shown in Figure S2f. As can be seen, the RIE process with the first recipe allows us to isotropically etch ZnSe/ZnMgSe material defining almost vertical sidewalls of the NPs. At the same time, $H_2$/Ar plasma also slightly etches the pre-defined HSQ SIL used as the hard mask. As can be seen in Figure S2f, the top part of the lens is truncated making the initial semi-spherical shape of the lens to be cylindrical.

The second dry etching recipe employed the mixture of $H_2$, Ar, and $CHF_3$ gases and was used to pattern NPs with spherical-shaped HSQ nanolenses on top. A typical image of the device etched with $H_2$/Ar/$CHF_3$ is shown in Figure S2g.

The performance of optical devices such as semiconductor SPSs is extremely sensitive to the quality of the device material and the number of defects. Therefore, after the dry etching step, wet-chemical polishing of the outer surface of the NPs was performed to remove the defects introduced by the RIE process as well as improve the smoothness of the sidewalls (see Figure S2d). The water-based solution of potassium dichromate (HBr:$K_2Cr_2O_7$:$H_2O$) was used as the etching solution. To perform a uniform polishing of NP structures, the chemical was stirred with a rotation speed of 300 rpm. The polishing of the samples was carried out for 5 s at room temperature and stopped by immersing the samples in the de-ionized water. As a result, the diameter of the pillars decreased by approximately 25 nm as can be seen in Figures S2f and S2g.

To subsequently enable a quantitative comparison between pillars with spherical shaped and cylindrical shaped HSQ lenses and to estimate the impact of the HSQ material on the photon collection efficiency, we removed HSQ lenses for a part of the fabricated NP emitters employing 1% hydrofluoric acid (HF). Finally, to prevent oxidation and thus degradation of the nanostructures, all the samples were immediately passivated with 10 nm $Al_2O_3$ oxide deposited using an atomic layer deposition system.

## 3. External quantum efficiency concept

For the development of practical and highly efficient single-photon sources, maximizing the external quantum efficiency is a highly important aspect. In particular, most of the reported single-photon sources emit photons isotropically which causes the issue of low collection efficiency of generated photons. To date, a couple of different approaches allowing to improve this factor for solid-state SPSs are developed and being pursued. Among those, the use of SIL that can be directly placed on top of a solid-state emitter is a common approach allowing to substantially improve the collection efficiency and thus enhance the total external quantum efficiency.[3,4] At the same time, the use of SIL which is made of low refractive material is an effective solution to overcome the total internal reflection problem[3] which is the origin of the low outcoupling efficiency of photons escaping from the emissive material.

### 3.1. Light outcoupling efficiency enhancement

The outcoupling efficiency ($\eta_{out}$) of photons through a flat surface is also limited by total internal reflection (TIR).[3] SILs remedy this problem to some extent by increasing the angle $\alpha_{TIR}$ under which TIR occurs and thus increasing the cone of light that can be collected through the surface. Figure S4a schematically shows a pillar covered with a spherical HSQ lens. Assuming that the point source with intensity $I_0$ emits isotropically in all spatial directions, the fraction of light that can be collected through the ZnMgSe surface is given by:

$$\eta_{out} = \frac{I(\alpha_{TIR})}{I_0} = \frac{1}{2} \int_0^{\alpha_{TIR}(n')} T(\alpha, n') \sin(\alpha)\, d\alpha$$

where

$$\alpha_{TIR}(n') = \arcsin n' \quad \text{with} \quad n' = \frac{n}{n_{ZnMgSe}}$$

is the angle of TIR at the interface of ZnMgSe and a medium with refractive index $n$ and $T(\alpha, n')$ is the transmission through this interface for light that is incident under an angle $\alpha$. The transmission can be calculated from the reflectivity $r(\alpha, n')$ using:

$$T(\alpha, n') = 1 - r^2(\alpha, n').$$

The reflectivity is given by the Fresnel equations and is different for TE or TM polarized light:

$$r_{TE}(\alpha, n') = \frac{\cos\alpha - \sqrt{n'^2 - \sin^2\alpha}}{\cos\alpha + \sqrt{n'^2 - \sin^2\alpha}} \quad \text{and} \quad r_{TM}(\alpha, n') = \frac{-n'^2\cos\alpha + \sqrt{n'^2 - \sin^2\alpha}}{n'^2\cos\alpha + \sqrt{n'^2 - \sin^2\alpha}}.$$

Using these equations, the outcoupling efficiency dependence on the refractive indices of the lens was calculated and is shown in Figure S4b. It can be seen that the efficiency can be increased by a factor of about 2.85 by covering the NP with the HSQ (SiO$_2$ for an approximation).

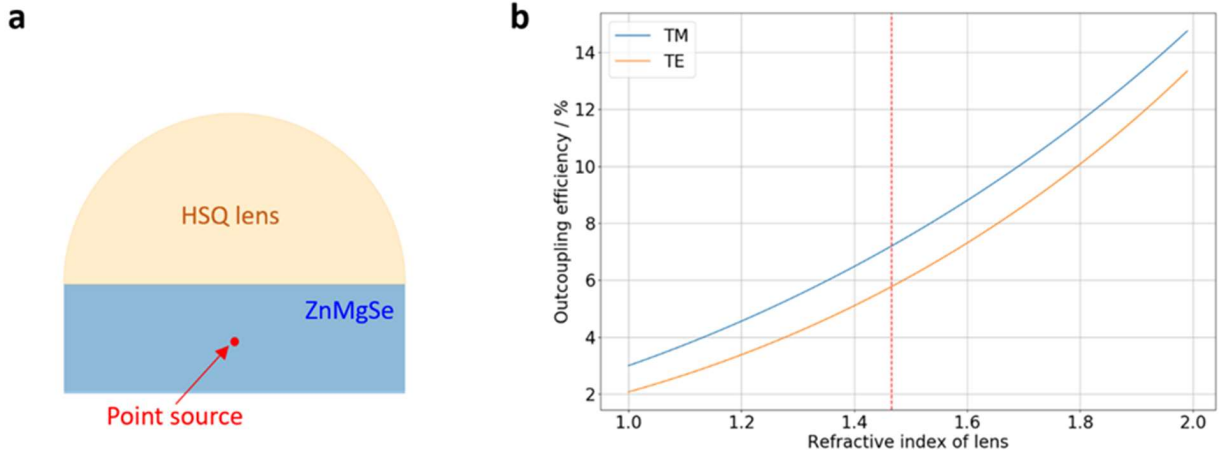

**Figure S4.** (a) Schematic drawing of a nanopillar with a spherical-shaped lens. (b) The outcoupling efficiency for TE and TM polarized light. The refractive index of SiO$_2$ is indicated by a red dashed line.

### 3.2. Light collection efficiency enhancement: Ray-tracing approach

HSQ lenses on top of ZnMgSe pillars help to increase also the light collection efficiency ($\eta_{coll}$) that can be estimated using a ray-tracing approach based on the following assumptions:

1) The emission of the donor is modeled using $N$ beams with an equal angular distance of 0.36° to each other that are emitted into the upper half-plane.

2) The path of each beam through the lens is traced using Snell's law:

$$n_1 \cdot \sin(\vartheta_1) = n_2 \cdot \sin(\vartheta_2)$$

where $n_1 = 1$ and $n_2 = 1.46655$ are the refractive indices of air and SiO$_2$, respectively and $\vartheta_1$ and $\vartheta_2$ are the angles under which the beam is incident onto the interface (concerning the surface normal) and under which the beam leaves the interface, respectively.

3) The beams are colored depending on whether they can be collected by the objective or not: If they can be collected by the objective, they are colored green. If they cannot be collected by the objective, either because they are completely reflected at the lens-air interface (i.e., TIR occurs) or because the angle under which they are emitted is larger than the acceptance angle of the objective (i.e. $\alpha_{crit} \approx 64.2°$), they are colored red.

4) The collection efficiency can be estimated by the fraction of beams that have been colored green divided by the total number N of rays:

$$\eta_{coll} = \frac{\#\ green\ beams}{N}$$

It is important to note, that the absolute values of the efficiency, calculated using this approach, do not represent the actual efficiency of which light can be collected from the pillar, as any losses at the ZnMgSe-lens interface are neglected and only emissions into the upper half-space were considered. For simplification, the transmission of the beams through the interface is assumed to be $T(\vartheta) = 1$ regardless of angle i.e. the beams are either completely transmitted or reflected (due to TIR). The ray-tracing approach can however give a rough quantification of the collection efficiency variation for instance if the donor is not in the center of the pillar. The simulation results are presented in Figures S4 and S5 for the pillars covered with spherical- and cylindrical-shaped lenses, respectively and are discussed below. The parameters that were used for the simulation are summarized in Table 1.

| Parameters | |
|---|---|
| NA of the objective | 0.9 |
| Total number of beams $N$ | 500 |
| Lens Radius $R$ | 125 nm |
| Range of donor position | [0 nm, 120 nm] |
| Range of spacer thickness (spherical) / Range of lens thickness (cylindrical) | [0 nm, 500 nm] |

**Table 1.** Parameters that were used for the ray-tracing simulation.

*Spherical-shaped lenses.* Figure S5a shows the dependence of the collection efficiency on the donor position as well as on the HSQ spacer thickness (i.e. HSQ thickness between the lens and ZnMgSe surface) for the pillar covered with a spherical-shaped HSQ mask. The simulation results demonstrate that this device yields an increased collection efficiency when an HSQ spacer of 40 - 110 nm thickness is introduced between the lens and ZnMgSe surface. Moreover, a strong increase in the collection efficiency is confirmed, the more the donor is located towards the center of the pillar.

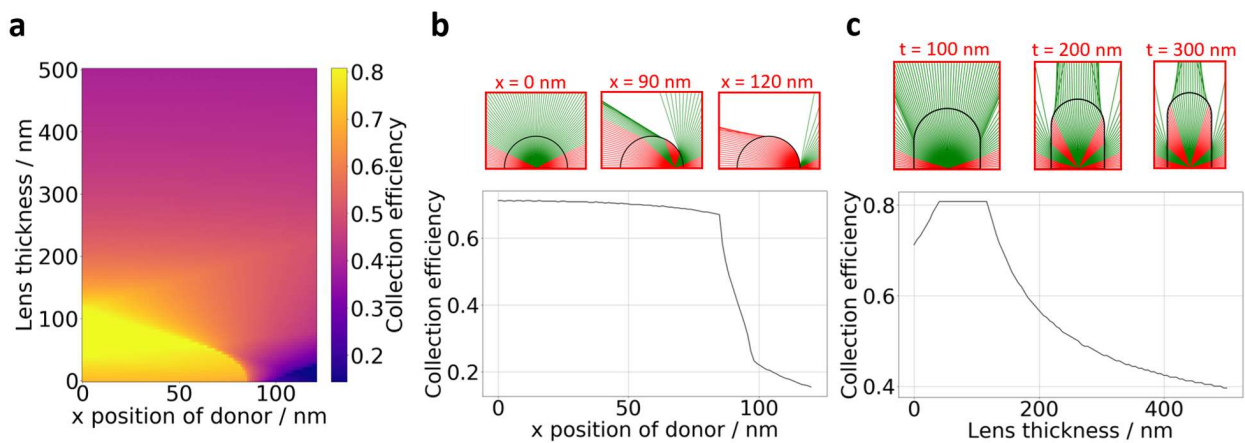

**Figure S5** Effect of the donor position on the collection efficiency for the pillars covered with spherical-shaped HSQ nanolenses. (a) Contour plot showing the change in the collection efficiency $\eta_{coll.}$ as a function of donor position and spacer layer thickness. (b) The change in the collection efficiency when the donor is moved from the center towards the NP edge. No spacer layer is considered here. The three upper insets show the results of the ray-tracing simulation in the cases where the donor has shifted by x = 0 nm, x = 90 nm, and x = 120 nm (from left to right) from the center. The color of the beams indicates whether they can be collected by the objective (green beams) or not (red beams). (c) Plot of the collection efficiency change as a function of the spacer thickness when the donor is located at the center of the pillar (x = 0 nm). The three upper insets (from left to right) show the results of the ray-tracing simulation for spacer thicknesses of t = 100 nm, t = 200 nm, and t = 300 nm, respectively.

Figure S5b shows the collection efficiency of the spherical lens depending on the position of the donor. It can be seen that the efficiency decreases by a factor of approximately 4.7 due to TIR when the donor is moved from the center of the pillar towards the edge (see insets in Figure S5b). Additionally, Figure S5b demonstrates that the beam is collimated with an increasing angle to the surface normal when the donor is moved further away from the center. This means that part of the light can't be collected anymore due to the limited aperture of the objective.

Figure S5c shows the effect of adding an HSQ spacer layer below the spherical lens. The collection efficiency can be maximized for the spacer layer thickness between 40 nm and 110 nm. In this case, the light is collimated towards the objective. Moreover, an additional portion of the light can be collected through the side flanks of the spacer (see insets in Figure S5c). If the spacer thickness is increased even further, the light is again reflected due to TIR at the side of the spacer and the lens. Hence, the collection efficiency again decreases with a further increase of the HSQ spacer.

Therefore, the simulation results show that three geometrical aspects maximize the light collection efficiency from the NPs: (1) adding a spherical-shaped nanolens on top of the pillar; (2) a spacer layer with a thickness of 40-100 nm between nanolens and ZnMgSe layer; (3) the donor is located close to the center of the pillar. If all three requirements are fulfilled, an enhancement factor of about 15 can be achieved compared to that without any HSQ coverage on top of the NPs.

*Cylindrical-shaped lenses.* Figure S6 demonstrates the simulation results for cylindrical lenses. The contour plot in Figure S6a shows that the collection efficiency is generally increased for lenses that are 120-160 nm thick. At the same time, in contrast to the spherical lenses, the optimum donor position is not always in the center, but for thin cylindrical lenses (t < 120 nm) the optimum is reached if the donor is closer to the edge of the lens. Figure S6b shows the change in the collection efficiency if the donor position is shifted along the x-axis (horizontally) in the case of a 125 nm thick cylindrical lens. We observe that the collection efficiency decreases gradually by a factor of approximately 1.7 as the donor is moved from the center toward the edge of the pillar.

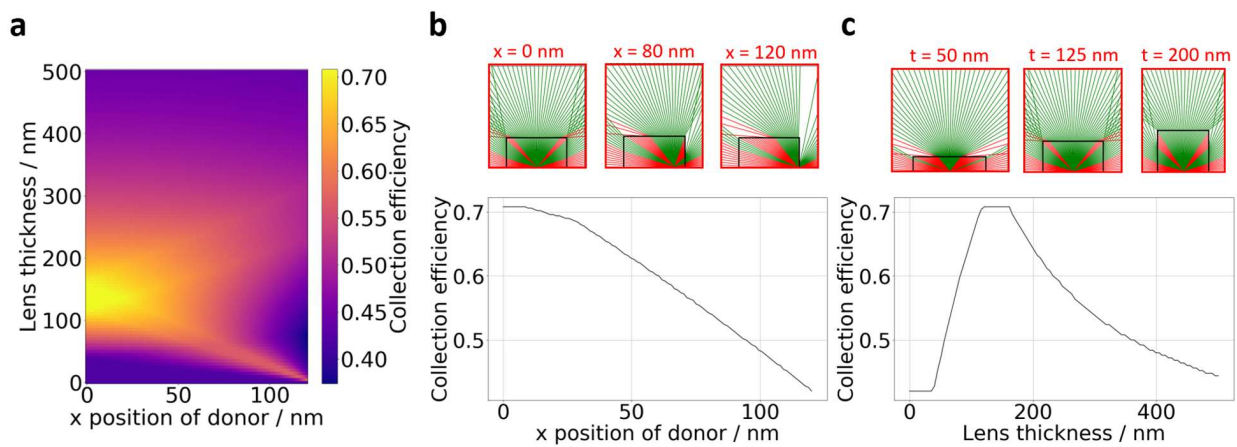

**Figure S6.** Effect of the donor position on the collection efficiency $\eta_{coll}$ for the pillars covered with cylindrical-shaped nanolenses (a) Contour plot showing the change in $\eta_{coll}$ as a function of the donor position and the lens thickness. (b) Collection efficiency changes as a function of the donor position for a cylindrical lens with a thickness of t = 125 nm. The three upper insets show the results of the ray-tracing simulation in the cases where the donor is shifted by x = 0 nm, x = 80 nm, and x = 120 nm (from left to right) from the center. (c) Dependence of the collection efficiency on the thickness of the cylindrical HSQ lens when the donor is located at the center of the lens (x = 0 nm). The three upper insets show (from left to right) the results of the ray-tracing simulation in the cases where the thickness of the cylindric lens is t = 50 nm, t = 125 nm, and t = 200 nm, respectively.

Figure S6c shows that similarly to the spherical lens, the collection efficiency increases up to a certain thickness of the cylindrical-shaped lens, as more light can be collected through the side flanks. If the thickness is further increased, TIR occurs at the side flanks and the intensity is reduced again. At the same time, the insets show that in contrast to the spherical-shaped lenses, the light is not collimated upwards. It can therefore be expected that using an objective with a smaller numerical aperture (NA), would reduce the efficiency of cylindrical lenses more severely compared to that of spherical-shaped lenses.

## 4. External quantum efficiency concept: Numerical simulations using Lumerical software

We use the Finite-Difference Time-Domain (FDTD) solver from the commercial software Lumerical to perform the numerical simulations of outcoupling efficiency for NPs covered with HSQ nanolenses. Figures S7a-c show a 3D simulation model used for the simulations, which consists of the GaAs substrate and an 85 nm thick ZnMgSe NP covered with a spherical HSQ nanolens (approximated as $SiO_2$). An electric dipole oriented in the horizontal direction (x or y) was employed to model the donor in ZnMgSe, which is vertically placed at the center of the ZnMgSe layer.

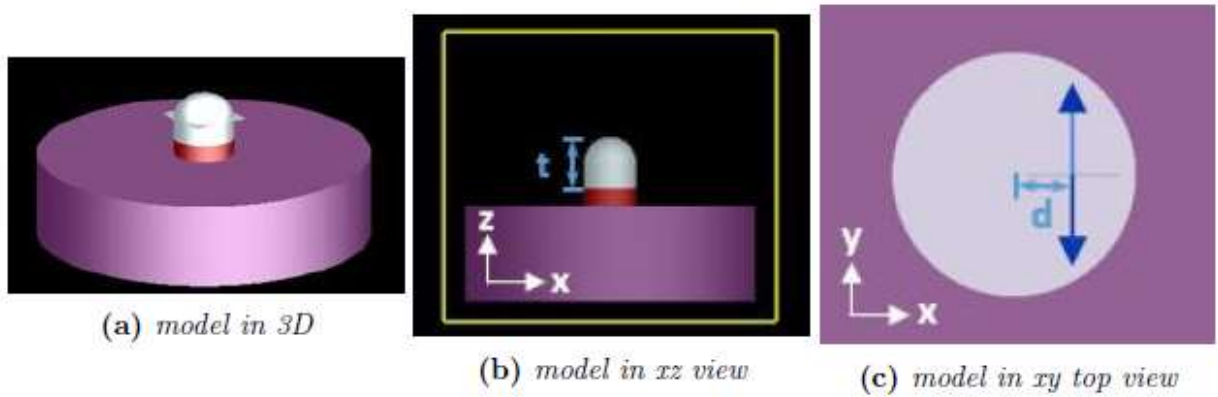

(a) model in 3D  (b) model in xz view  (c) model in xy top view

**Figure S7.** (a) 3D model of a nanopillar with a spherical-shaped HSQ mask with a cylindrical spacer. (b) 2D view of the model enclosed by a box of field monitors in the air. (c) A top view of the model with a dipole in y-direction laterally displaced from the center with a distance d.

*HSQ spacer effect.* To investigate the influence of HSQ thickness, we varied the spacer thickness from 0 nm to 500 nm in a step of 25 nm, for both the spherical-shaped HSQ mask and the cylindrical one. The dipole in the y-direction is placed laterally in the center of the NP and the simulation with the x-direction dipole is omitted due to symmetry reasons. With the closed box of monitors in the air enclosing the whole NP structure, the local electromagnetic fields on the box surfaces are recorded and the far-field radiation pattern is therefrom calculated.

Figures S8a and S8b show the dependence of the collected emission power on HSQ spacer thickness for NPs with the spherical-shaped HSQ and cylindrical HSQ lenses, respectively. The collected emission power with different values of numerical aperture (NA) is calculated by integrating over the angular cone defined by NA in the far-field radiation pattern. Figure S8c specifies the case of NA=0.9. The collected emission power is normalized to the total emission power of the same dipole when it is placed in a homogeneous ZnMgSe medium without any other material or structures. A non-monotonic behavior of the collected emission power with the increase of the HSQ spacer thickness is observed for both types of lenses. However, as can be seen in Figure S8c, up to 135% efficiency enhancement can be achieved for the pillars covered with 125 nm thick HSQ as compared to the pillars without any lenses, which is in good agreement with experimentally observed results.

The far-field radiation patterns with the spherical-shaped HSQ lenses at different HSQ thicknesses are shown in Figure S9. We see that the emission becomes more directional with increasing HSQ thickness.

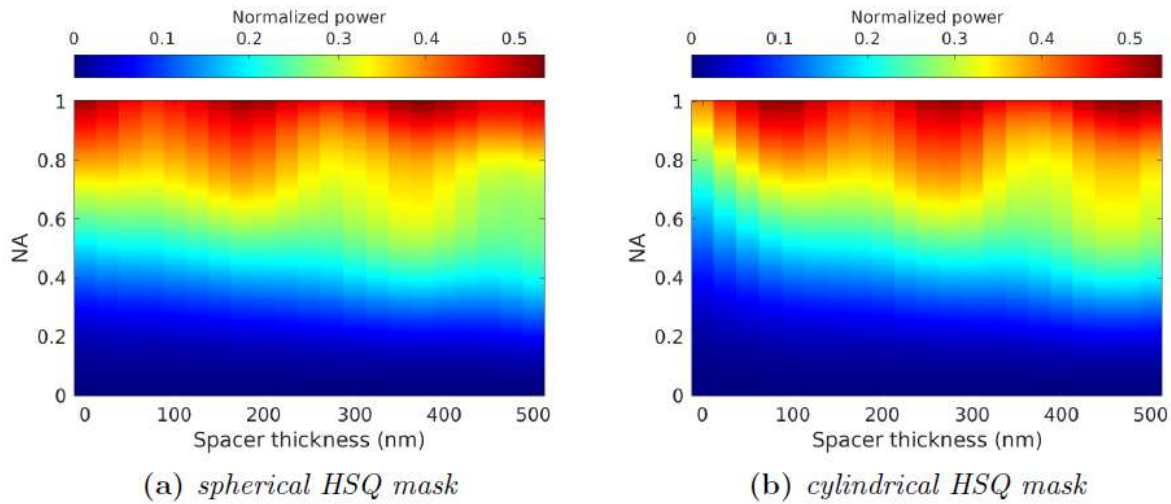

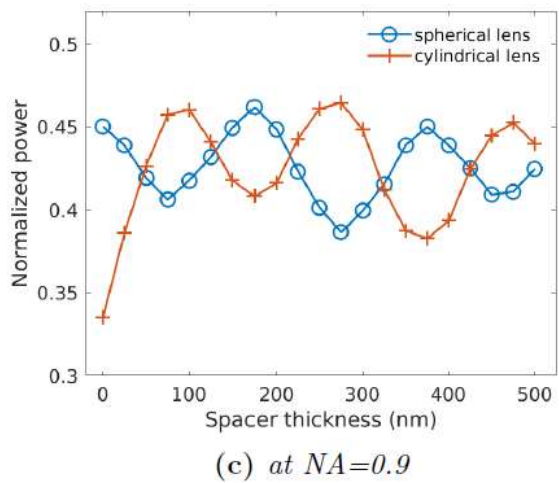

**Figure S8.** Emission power collected with different NA and varying HSQ thicknesses for (a) spherical-shaped and (b) cylindrical HSQ lenses. (c) Dependence of the emission power versus the HSQ thickness obtained for NA=0.9.

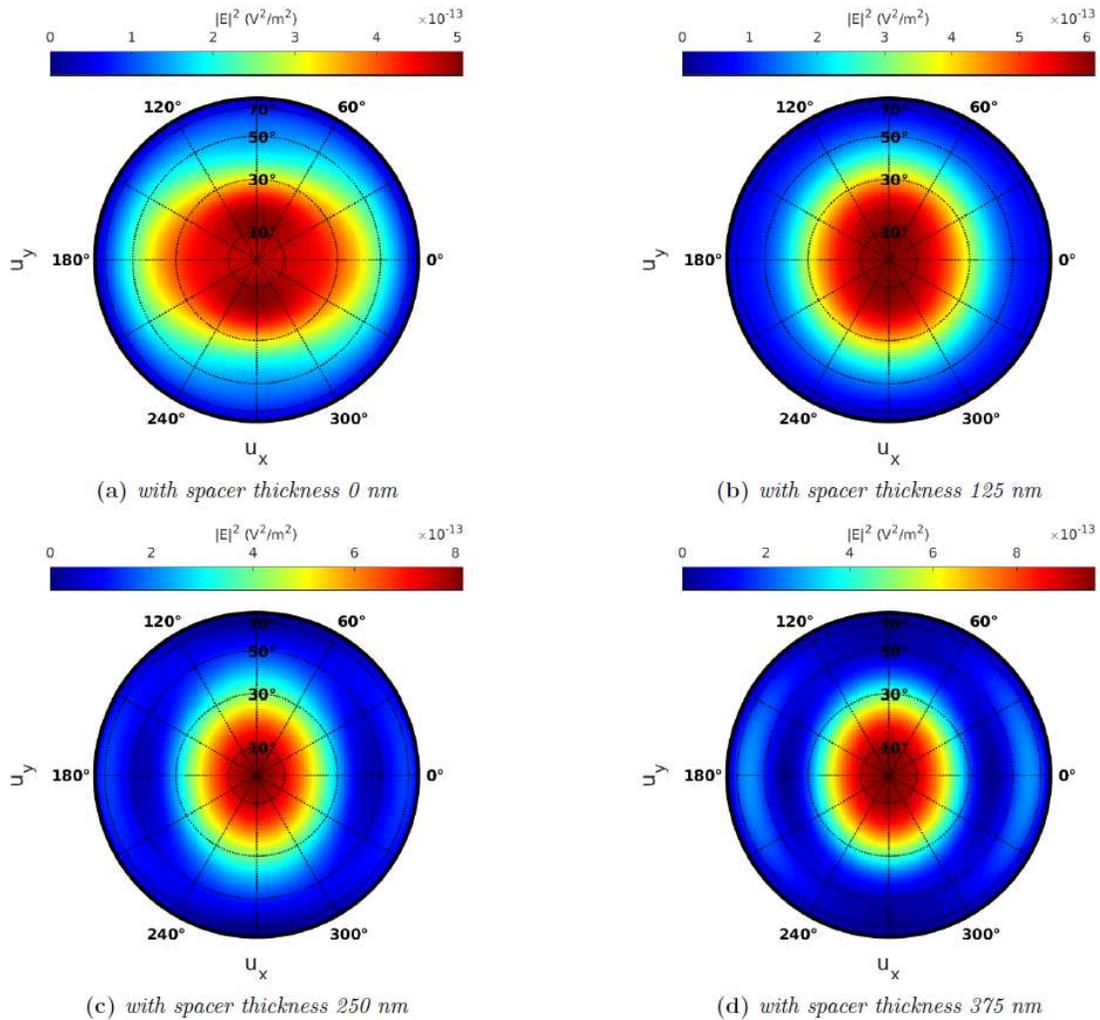

**Figure S9.** Far-field radiation patterns for spherical-shaped HSQ lenses with different HSQ spacer layer thicknesses.

*Effect of the donor position.* To investigate the influence of the lateral position of the donor, we varied the x position of the dipole from the center to 120 nm towards the NP edge (with a step size of 15 nm). This series of simulations are performed for three different cases of HSQ mask: 125 nm of spherical HSQ mask (without spacer), 125 nm of cylindrical HSQ mask, and without any HSQ mask on top (see Figures S10a, S10b, and S10c, respectively). The simulations are carried out for x-direction and y-direction dipoles respectively and the final simulation results are averaged out.

Figure S10d shows the dependence of the collected emission power on the dipole position for the case of NA=0.9. As can be seen from Figure S10d, the collected power decreases for both types of lenses as the dipole is moved towards the edge of the NP. Interestingly, for the spherical and cylindrical HSQ lenses, the decrease in the collected power by approximately a factor of 8 is observed, while for the case without any lens, the reduction factor is about 6. It should be noted that the obtained results are also consistent with the experimental data presented in the paper. The far-field radiation patterns for spherical HSQ mask varying dipole x-position are shown in Figure S11. The results reflect the strong impact of the donor position on the external quantum efficiency for the developed single-photon emitters.

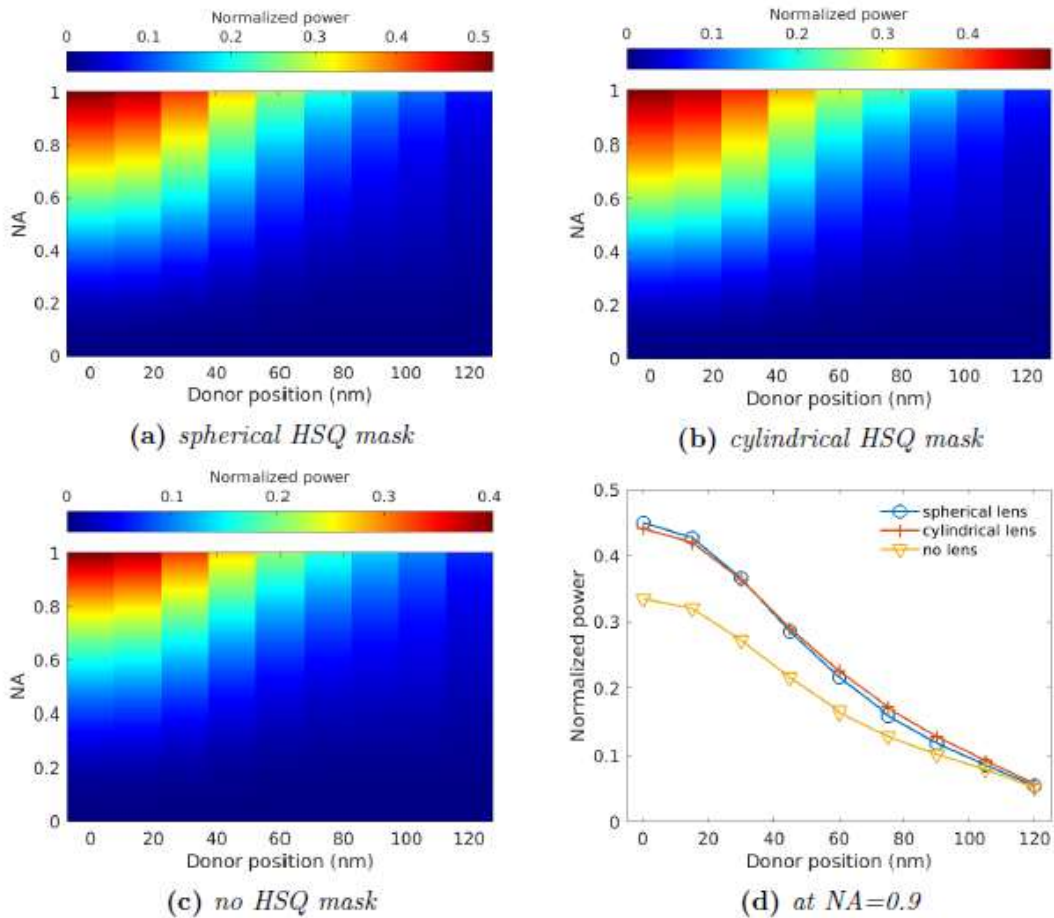

**Figure S10**. Emission power collected with different NA when laterally varying the dipole positions (average of dipoles in x- and y-direction) from the center to the edge.

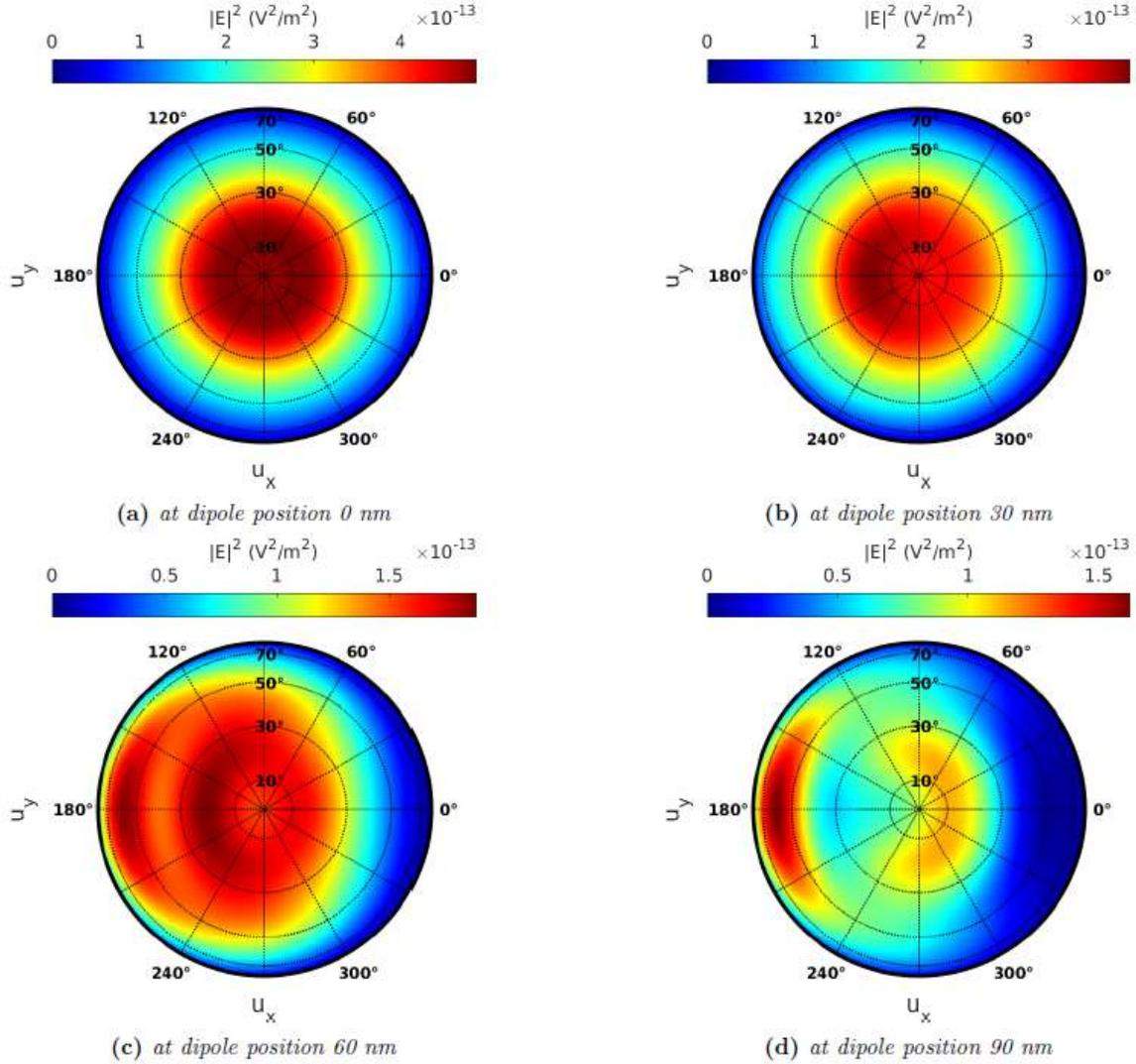

**Figure S11.** Far-field radiation patterns for the spherical-shaped HSQ lens when varying the lateral dipole position (average of dipoles in x- and y-direction) from the center to the edge of the NP.

### 5. Calculated band diagram of ZnMgSe/ZnSe/ZnMgSe QW structure

Figure S12 shows the calculated position of the conduction and valence bands and the component of the wave functions in the direction perpendicular to the layer growth. The slight bending of the bands is due to the electrical field associated with the ionized Cl atoms present in the ZnSe QW. Due to the relatively low Mg content (between 10% and 13%) in the ZnMgSe barriers, the band offset of the strained type – I ZnSe/MgZnSe heterostructure is below 132 meV for the conduction and 41 for the valence band (heavy holes), correspondingly. This leads to the formation of not more than two-electron and heavy-hole subbands in the ZnSe QW with a width of less than 5 nm.

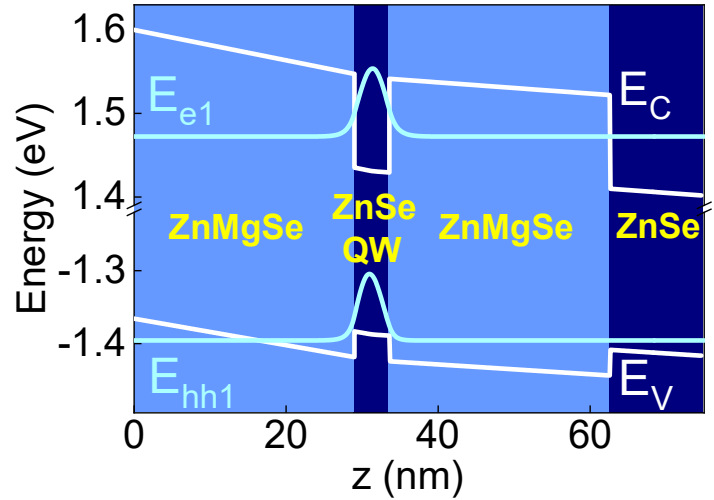

**Figure S12**. Band diagram and the squared electron and heavy hole eigenfunctions calculated for a typical ZnSe/ZnMgSe sample with a 4.7 nm thin ZnSe QW, 11% Mg concentration in the ZnMgSe barriers, and $1.4 \times 10^{10}$ cm$^{-2}$ Cl doping concentration at the middle of the QW.

# REFERENCES FOR THE SUPPORTING INFORMATION